\documentclass[prb]{revtex4}
\usepackage{feynmp}%%%{feynmp}self-energy
\usepackage{graphicx}
\begin{document}
\begin{fmffile}{fmfmy}
\title{Conservation laws in the quantum Hall Liouvillian theory and its generalizations}
\author{Joel E. Moore}
\affiliation{Department of Physics, University of California, Berkeley, CA 94720; \\
Materials Sciences Division,
Lawrence Berkeley National Laboratory, Berkeley, CA 94720}
\begin{abstract}
It is known that the localization length scaling of noninteracting electrons near the quantum Hall plateau transition can be described in a theory of the bosonic
density operators, with no reference to the underlying fermions.
The resulting ``Liouvillian'' theory has a $U(1|1)$ global supersymmetry
as well as a hierarchy of geometric conservation laws related to
the noncommutative geometry of the lowest Landau level (LLL).
Approximations to the Liouvillian theory contain quite different physics from standard
approximations to the underlying fermionic theory.  Mean-field and large-$N$
generalizations of the Liouvillian are shown to describe problems of noninteracting bosons that enlarge the $U(1|1)$ supersymmetry to $U(1|1) \times SO(N)$ or $U(1|1) \times SU(N)$.

These noninteracting bosonic problems are studied numerically for $2 \leq N \leq 8$ by Monte Carlo simulation and compared to the original $N=1$ Liouvillian theory.
The $N>1$ generalizations preserve the first two of the hierarchy of geometric conservation laws, leading to logarithmic corrections at order $1/N$ to the diffusive large-$N$ limit, but do not preserve the remaining conservation laws.  The emergence of nontrivial scaling at the plateau transition, in the Liouvillian approach, is shown to depend sensitively on the unusual geometry of Landau levels.
\end{abstract}
\maketitle

\section{Introduction}
The original work of Anderson on localization in noninteracting electronic systems~\cite{anderson}
proposes a simple test for the existence of extended states in a disordered system.
When an electron is added at the origin at $t=0$, its mean squared displacement
is finite for all times if all states are localized:
$\langle \langle R^2(t) \rangle \rangle \leq C$ for some constant $C$.  If there are extended states in the system, then the electronic motion is diffusive for long times, and $\langle \langle R^2(t) \rangle \rangle \sim D t$.  These two alternatives are not exhaustive, however.  The motion of lowest-Landau-level electrons in a white-noise potential, starting at $t=0$ from a localized wave packet, satisfies $\langle \langle R^2(t) \rangle \rangle \sim C t^\alpha$ for $\alpha \approx 0.79$, and numerics suggest that this exponent $\alpha$ is a general property of the universality class of the quantum Hall plateau transition.  The physical origin of this exponent is the ``two-parameter scaling'' 
that describes the Hall transition: both the diagonal and transverse conductivities are needed to characterize the system~\cite{khmelnitskii}, unlike in the zero-field case.

Note that the universal quantity $\alpha$ can be defined without any reference to the
electron {\it energy} or to the electron {\it wavefunction}.  The electron density correlations are already sufficient to obtain $\alpha$, which is related to the localization length exponent $\nu$ by $\alpha = 1-{1 \over 2\nu}$.  This insight is the basis of the Liouvillian approach to the plateau transition introduced by Sinova, Meden, and Girvin~\cite{smg}.  An exact theory which contains $\alpha$ can be written down in terms of the bosonic electron density operators, and while this theory has not yet been solved, it has been studied analytically in several different limits~\cite{gurarie,boldyrev,mooresinova}.  Much of the interest in such noninteracting models stems from the fact that the value of $\nu$ from such models is consistent with that found in some experimental samples~\cite{wei}.   Adding short-ranged interactions does not modify this value~\cite{dhlee}, but the effects of long-ranged interactions are currently unclear.

This paper discusses the physical origin of critical scaling and anomalous dimensions in the Liouvillian picture.  The bosonic density operators $\rho_q$ in the
lowest Landau level satisfy a linear first-order evolution equation; averaging this evolution equation over disorder cannot be done analytically.  However, previous work has considered mean-field and large-$N$ limits where the disorder-averaged theories can be studied analytically (here $N$ is the number of density ``flavors'', and $N=1$ is the
physical case).  Part of this paper considers such theories in more detail and shows how
the large-$N$ theories represent theories of noninteracting bosons, which can be studied
using Monte Carlo simulations just as for $N=1$.

An important question about the Liouvillian approach is how the seemingly simple Lagrangian of the density operators can show the nontrivial $\alpha<1$ scaling behavior observed numerically.  We argue that long-lived fluctuations resulting from a hierarchy of conservation laws in the Liouvillian formulation cause the scaling.  This picture is supported by analytical and numerical study of mean-field and large-$N$ approximations which preserve some but not all of these conservation laws.  Previous work found diffusive behavior ($\alpha = 1$) for the mean-field and large-$N$ limits, with a diverging set of corrections at order $1/N$.  These logarithmic corrections occur because of the first nontrivial conservation law beyond energy and particle number, that of ``chop'',
a measure of the variation in the particle density.  Note that these logarithmic corrections occur for a different physical reason than the logarithmic corrections which appear in the
theory of weak localization~\cite{leeramakrishnan}.

There is a hierarchy of higher conservation laws in the $N=1$ problem which are not preserved in the $N>1$ generalizations; these higher conservation laws account for the differences we find numerically between $N=1$ and $N>1$.  These conservation laws can be expressed compactly in terms of electron translation around polygonal
paths.  They essentially reflect the nonindependence of the density operators in an LLL state: there are many more density operators than independent LLL states, so these density operators must be related to each other.  (Conservation laws in the Liouvillian formalism are also discussed in~\cite{boldyrev}, but those explicitly involve the disorder potential and hence differ from the geometric conservation laws discussed here.)

The standard field-theoretic description of the quantum Hall plateau transition is as a nonlinear sigma model with topological term~\cite{pruisken}.  This
description has generated a great deal of subsequent work and clarified how the plateau transition fits into the general description of noninteracting localization problems in terms of nonlinear sigma models (NL$\sigma$Ms).  However, it has so far been difficult to find a clear understanding of how the topological term yields power-law scaling in models with noncompact target spaces, and efforts to obtain scaling via e.g. instanton gas calculations are difficult to confirm numerically.  Supersymmetric nonlinear sigma models into compact spaces in two dimensions can be understood quite thoroughly~\cite{readsaleur}.
These difficulties motivate work on other approaches such as the Liouvillian, which has
already yielded an improved understanding~\cite{gurarie} of the connection between the quantum Hall transition and classical percolation~\cite{trugman}.

Now we explain the connection~\cite{smg} between the localization length exponent $\nu$ and the power-law spreading of a wave packet discussed above.
The critical exponent $\nu$ describes the divergence of the localization
length near the critical energy:
\begin{equation}
\xi(E) = \xi_0 \left({E_c \over E - E_c}\right)^\nu.
\end{equation}
An electron added in a ``typical'' localized state (not an eigenstate) will project
onto states of many different energies.  The density of states is believed to be nonsingular at the critical energy, so it is reasonable to assume that on average the added electron
projects equally onto states of different energies.

The second assumption needed is that
electron motion is diffusive in each state up to the localization length at that
energy, then stops:
\begin{equation}
\langle \langle R^2(t) \rangle \rangle_E = \cases{Dt&if $Dt < \xi(E)^2$\cr\xi(E)^2&
if $Dt \geq \xi(E)^2$}.
\end{equation}
Here $D$ is a diffusion constant with a nonsingular dependence on energy.
Then the mean squared displacement, averaged over energy, is
\begin{equation}
\langle \langle R^2(t) \rangle \rangle = (D t)^{1 - {1 \over 2 \nu}} {\xi_0}^{1 \over \nu}
\propto t^\alpha.
\end{equation}
This prediction $\alpha = 1 - {1 \over 2 \nu}$ can be verified numerically, as discussed in section V.

The many conservation laws for the density operators mentioned above explain several previous results on mean-field and large-$N$ theories, and suggest a picture which is confirmed below by Monte Carlo numerics.  A mean-field theory gives only diffusive spreading of a wavepacket at long times~\cite{smg}.  A slight modification to this mean-field theory allows it to be interpreted as a large-$N$ limit, where $N$ is the number of ``flavors'' of density operators.  Then the leading corrections in $1/N$ include an infinite series of maximally crossed diagrams, whose sum diverges even though each individual diagram is finite.  The first extra conservation law, of ``chop'', explains this divergence in the same way that particle number conservation explains the singularity in weak localization.

Part of this paper shows how the $N>1$ generalizations can be represented and studied numerically as systems of noninteracting bosons in random potentials.  The density evolution for finite systems for $2 \leq N \leq 8$, found via Monte Carlo simulations, does interpolate between the analytic result for $N=\infty$ and the physical $N=1$ quantum Hall problem.  However, higher conservation laws at $N=1$ do not have equivalents for $N>1$, and as a consequence the logarithmic corrections at order $1/N$ seem unlikely to indicate a power law for $N>1$.  The numerical density evolution for the largest systems studied at $N=2$, which correspond to Liouvillian matrices of size approximately $10^5 \times 10^5$, is found to behave nearly diffusively for long times, suggesting that
the true density evolution for$N>1$ in the thermodynamic limit is diffusive (possibly with logarithmic corrections), rather than scaling with a power law slower than diffusion as
for the physical case $N=1$.

An important difference between the quantum Hall transition and the superficially
similar classical percolation transition is that the former seems numerically to have only one nontrivial exponent in the spreading of the particle distribution function, while the latter is ``multifractal'' and has different exponents for different moments of the particle distribution function~\cite{mooreranlev}.  Nevertheless, some aspects of the quantum-mechanical spin quantum Hall transition are exactly described by sums over classical percolation hulls~\cite{gruzberg}.  One known property of the Liouvillian action (which is derived in the following section) is that truncating higher terms in its nonlocal interaction to make it a local theory leads to the classical percolation hull problem~\cite{gurarie}: the full quantum Hall problem is essentially the unique generalization of this classical problem to a noncommutative space.  Most applications of noncommutative geometry to the quantum Hall effect have focused on the plateaus~\cite{sondhi,fradkin}, which are currently less opaque than the transitions; the integer transition is a major open problem in condensed matter theory which has a simple statement in terms of a field theory in noncommutative geometry.  In order to be accessible by both condensed matter and high-energy physicists, this paper attempts to be self-contained and reviews some previous work in sections II and III.

The outline of this paper is as follows.  Section II reviews the Liouvillian theory set up to calculate the density correlation functions. Section III contains analytical results on this theory and its generalizations to multiple flavors of density operators.  The density operators for a general LLL state are shown in Section IV to satisfy a hierarchy of geometric constraints, which in turn implies that the evolution of density operators under the Liouvillian must preserve a number of conservation laws.  The large-$N$ generalizations of the Liouvillian preserve the first two conservation laws, of particle number and chop, but not the higher conservation laws.  Section V uses Monte Carlo simulations of finite-size realizations of the Liouvillian and its $N>1$ generalizations to test the analytic picture from previous sections, and ends with a summary of the main conclusions.

\section{Review of the Liouvillian approach}
\label{review}

The electron density of a single-particle state in the lowest Landau level has a
simple time evolution which is {\it closed}: knowledge of the density at one
time determines it for all later times.  The Liouvillian equation of motion for the density operators in a single disorder realization is derived in the first part of this section.
Later sections of this paper are chiefly concerned with efforts to average this equation over disorder, either numerically or analytically, and extract information about universal quantities such as $\nu$.

The noninteracting Hamiltonian of a single 2D electron in a constant magnetic field
$B {\bf \hat z}$ and random potential $v$ is, after
projecting to the lowest Landau level (LLL),
\begin{equation}
H = \sum_{\bf q} v(-{\bf q}) {\hat \rho}_{\bf q}.
\label{startham}
\end{equation}
Throughout this paper, we will assume the strong-field limit and ignore complications
from spin and Landau-level mixing.  The random potential components are independent
Gaussian variables:
\begin{eqnarray}
v(\bf q) &=& v({\bf -q})^*\cr
\langle \langle v({\bf q}) \rangle \rangle &=& 0 \cr
\langle \langle v({\bf q}) v({\bf q^\prime})\rangle \rangle &=& 
{2 \pi v^2 \over L^2} \delta_{{\bf q}+{\bf q^\prime}}
\end{eqnarray}
The system is on a square of side $L$ with periodic boundary conditions.  Each component of ${\bf q}$ takes on $N_\phi$ values, where $N_\phi = L^2 / (2 \pi \ell^2)$ is the number of flux quanta through the torus and $\ell=\sqrt{\hbar c/eB}$ is the magnetic length.

The equation of motion for the LLL-projected density operator ${\hat \rho}_{\bf q}$ is
\begin{equation}
{d \over dt} {\hat \rho}_{\bf q} = [H, \rho_{\bf q}] = \sum_{\bf q^\prime} v(-{\bf q^\prime})
[\rho_{\bf q^\prime},\rho_{\bf q}].
\label{seom}
\end{equation}
This commutator turns out to take an especially simple form because of the special
geometry of the lowest Landau level.

There is an identity which connects the LLL-projected densities $\rho_{\bf q}$ to
the ``magnetic translation'' operators $\tau_{\bf q}$:~\cite{girvinjach}
\begin{equation}
{\bar \rho}_{\bf q} = e^{-q^2 \ell^2/4} \tau_{\bf q}.
\end{equation}
The magnetic translation operator $\tau_{\bf q}$ translate a given LLL state by
$\ell^2 {\bf q} \times {\hat z}$.  The commutation relation of the $\tau_{\bf q}$ is fixed by the requirement that translation around a closed path pick up an Aharonov-Bohm phase
from the flux through the path:
\begin{equation}
[\tau_{\bf p}, \tau_{\bf q}] = 2 i
\sin(\frac{\ell^2 {\bf p} \wedge {\bf q}}{2}) \tau_{{\bf p}+{\bf q}}.
\end{equation}
Using this commutation relation and the Hamiltonian (\ref{startham}), the equation of motion for the $\tau_{\bf q}$ is
\begin{equation}
{\dot \tau}_{\bf q} = - i {\cal L}_{\bf q q^\prime}
\tau_{\bf q^\prime},
\label{eom}
\end{equation}
where the Liouvillian ${\cal L}$ is
\begin{equation}
{\cal L}_{\bf q q^\prime} = {2 i \over \hbar} v({\bf q} - {\bf q}^\prime)
\sin(\frac{\ell^2 {\bf q} \wedge {\bf q}^\prime}{2})
e^{-\frac{\ell^2}{4} |{\bf q} - {\bf q}^\prime|^2}.
\label{liouvillian}
\end{equation}

Now this evolution of the density and translation operators can be connected to
quantities of direct interest such as the disorder-averaged density correlation function.  The solution to the equation of motion (\ref{eom}) is
\begin{equation}
\tau_{\bf q}(t) = \sum_{\bf q^\prime} \left(e^{-i {\cal L} t}\right)_{\bf q q^\prime}
\tau_{\bf q^\prime}(0).
\label{lsolve}
\end{equation}
Let the averaged density correlation function be defined as
\begin{equation}
{\hat \Pi}({\bf q},t) \equiv -i {\theta(t) \over N \hbar \ell^2}
\langle \langle {\rm Tr}\{{\hat \rho}_{\bf q}(t) {\hat \rho}_{-{\bf q}}(0)\}\rangle\rangle.
\end{equation}
This is the energy integral of the usual energy-resolved density correlation
function~\cite{smg}.  Its significance will become clear in a moment.  The solution
(\ref{lsolve}) of the Liouvillian equation of motion then gives
\begin{equation}
{\hat \Pi}({\bf q},t) \equiv -i {\theta(t) \over \hbar \ell^2} e^{-\ell^2 q^2/2}
\langle \langle \left( e^{-i {\cal L} t} \right)_{\bf q q}\rangle\rangle.
\end{equation}

In order to understand how ${\hat \Pi}({\bf q},t)$ is connected to universal quantities
such as $\nu$, suppose that the
electron starts at $t=0$ in a localized state with given densities
${\hat \rho}_{\bf q}(0) = f_0({\bf q})$.
The function ${\hat \Pi}({\bf q},t)$ explains how the initial condition $f_0({\bf q})$ evolves over time, on average:
\begin{equation}
\langle \langle {\hat \rho}_{\bf q}(t) \rangle \rangle = {\hat \Pi}({\bf q},t) f_0({\bf q}).
\label{piest}
\end{equation}
Later this equation will be used to estimate ${\hat \Pi}({\bf q},t)$ numerically.
Now the density operators near $q=0$ can be used to extract moments of the density
distribution in real space: for example,
the squared displacement $\langle R^2 \rangle = \langle x^2 + y^2 \rangle$ at $t=0$ is given by, in the continuum limit, 
\begin{equation}
\langle R^2 \rangle = -\left({\partial^2 \over \partial^2 q_x} + {\partial^2 \over \partial^2 q_y}\right) f_0({\bf q}).
\label{characteristic}
\end{equation}

For a localized initial state of one electron, $f_0({\bf q}) \rightarrow 1$ as
$q \rightarrow 0$.  At long times $t$, we expect power-law spreading of the wave packet: $\langle \langle R^2(t) \rangle \rangle \sim t^{1 - {1 \over 2 \nu}}$.  The corresponding
statement in terms of ${\hat \Pi}({\bf q},t)$ is
\begin{equation}
-\left({\partial^2 \over \partial^2 q_x} + {\partial^2 \over \partial^2 q_y}\right)
\Pi({\bf q},t) \propto t^{1 - {1 \over 2 \nu}}.
\end{equation}
Hence the universal quantity $\nu$ is contained in the long-time, small-momentum behavior of ${\hat \Pi}({\bf q},t)$.
For simple diffusion, we would have had ${\hat \Pi}({\bf q},t) = e^{-D q^2 t}$,
and $\langle \langle R^2 \rangle \rangle \sim D t$, as expected.

%mention scaling form of $\Pi$
%mention percolation, how this is an exact rather than effective theory, noncommutative
%geometry.

%\section{Percolation limit and noncommutative geometry}

%WKB picture (A. Zee)

\section{Analytic results}

The preceding section reviewed how the critical exponent $\nu$ of the plateau
transition, a problem of noninteracting fermions, is contained in the Liouvillian theory of the {\it bosonic} density operators.  One would like to understand how the special
features of the plateau transition are reflected in the Liouvillian and cause the bosonic
fields to behave in such a complex manner.  The first part of this section reviews the mean-field and large-$N$ theories of the Liouvillian.  The supersymmetry technique (reviewed in~\cite{efetov}) is used to obtain a field-theoretic
representation of the Liouvillian~\cite{smg} with $U(1|1)$ supersymmetry.

The mean-field or large-$N$ limit of the Liouvillian shows diffusion of the
density.  Corrections to this limit are therefore localizing, i.e., reduce the exponent $\alpha$ in $\langle \langle R^2(t) \rangle \rangle \sim t^{\alpha}$ to its physical
value $\alpha \approx 0.79$.  The $1/N$ correction is examined analytically for
evidence of this localizing effect, and the $N>1$ theories are studied numerically in
the following section.
As described in more detail in section V, the orthogonal large-$N$
theory~\cite{mooresinova} corresponds to $N$ flavors of bosonic density operators, not to $N$ flavors of (fermionic) electrons; it is hence inequivalent to e.g. the $N$-orbital theory of Oppermann and Wegner~\cite{oppermann}.  The theory also has some unusual features characteristic of noncommutative-geometric theories, e.g., perturbation theory can be organized in such a way that every individual diagram is convergent, without any external regularization.

Some analytical results are obtained for a large-$N$ generalization where the symmetry of the theory is enlarged to $U(1|1) \times SO(N)$.  There are actually many different possible large-$N$ generalizations, some of which are equivalent to the original $N=1$ theory.  We explain in detail the different physical situations corresponding to
the orthogonal generalization $U(1|1) \times SO(N)$, a unitary generalization
$U(1|1) \times SU(N)$, and the most symmetric generalization $U(N|N)$, which is
actually equivalent to the original $N=1$ theory with $U(1|1)$ symmetry.  All these supersymmetric theories will become much more concrete in the following section, when they are realized as different conditions on
the random potentials of a problem of $N$ flavors of noninteracting bosons.

The second part of this section shows
how the quantum Hall dynamics cause the disorder-averaged correlation functions
to obey a set of geometric identities.  The simplest of these is related to the conservation of ``chop'', the variation in the density for states in the LLL.  This conservation law is important for understanding how scaling appears at the transition beyond mean-field theory.  The soft mode which appears at order $1/N$ in the large-$N$ theory can be
shown to exist beyond perturbation theory as a result of this conservation law.

We begin by reviewing the self-consistent Born approximation to $\Pi(q,\omega)$, which
will later be justified as a large-$N$ limit.  The disorder-averaged density propagator $\Pi(q,\omega)$ is represented in perturbation theory to $n$th order in the potential strength $v^2$ by diagrams with a single density line and $n$ interaction lines.  The
propagator of the magnetic translation operator has dimensions of time and is defined
by
\begin{equation}
{\hat \Pi}(q,\omega)\equiv \hbar \ell^2 e^{l^2 q^2 / 2} \Pi(q,\omega).
\end{equation}
The Feynman rules for calculation of this quantity follow from averaging the disorder potentials in the Liouvillian (\ref{liouvillian}).
The bare electron line is $(\omega+i\delta)^{-1}$, and the Liouvillian interaction vertex induced by averaging over disorder has an unusual form:
\begin{eqnarray}
\langle \langle {\cal L}_{{\bf p},{\bf p}+{\bf k}}
{\cal L}_{{\bf q}+{\bf k},{\bf q}} \rangle \rangle &=&
{4 \pi v^2 \over \hbar^2 L^2} e^{-\ell^2 k^2 / 2} 
\sin(\frac{\ell^2}{2} {\bf p} \wedge {\bf k}) \sin(\frac{\ell^2}{2}{\bf q}\wedge {\bf k})\cr
&=&\begin{fmfchar*}(40,15)
\fmfstraight
\fmfbottom{di,z0,z2,z3,z1,do}
\fmf{plain,label=${\bf p}\vphantom{+{\bf k}}$}{di,z0}
\fmf{plain,label=${\bf q}+{\bf k}$}{z3,z1}
\fmf{plain,label=${\bf q}\vphantom{+{\bf k}}$}{z1,do}
\fmf{plain,label=${\bf p}+{\bf k}$}{z0,z2}
\fmf{dashes,left}{z0,z1}
%+{\bf k}
\fmfdot{z0,z1}
\end{fmfchar*}.
\end{eqnarray}

Now we use the above rules to write down the self-consistent Born
approximation (SCBA)~\cite{smg}, which sums all the diagrams with no crossing of interaction lines.  In a moment we will show how a very similar approximation emerges
as the exact large-$N$ limit of the Liouvillian.  Using double lines to denote the SCBA
propagator, we have
\begin{eqnarray}
\begin{fmfchar*}(20,10)
\fmfstraight
\fmfbottom{di,do}
\fmf{dbl_plain,label=${\bf q}$}{di,do}
\end{fmfchar*}
&=&
\begin{fmfchar*}(20,10)
\fmfstraight
\fmfbottom{di,do}
\fmf{plain,label=${\bf q}$}{di,do}
\end{fmfchar*}
+
\begin{fmfchar*}(20,10)
\fmfstraight
\fmfbottom{di,z0,z1,do}
\fmf{plain,label=${\bf q}$}{di,z0}
\fmf{plain,label=${\bf q}$}{z1,do}
\fmf{plain,label=${\bf q}_1$}{z0,z1}
\fmf{dashes,left}{z0,z1}
\fmfdot{z0,z1}
\end{fmfchar*}
+\ldots ({\rm all\ noncrossing})\cr
&=&
\begin{fmfchar*}(20,10)
\fmfstraight
\fmfbottom{di,do}
\fmf{plain,label=${\bf q}$}{di,do}
\end{fmfchar*}
+
\begin{fmfchar*}(20,10)
\fmfstraight
\fmfbottom{di,z0,z1,do}
\fmf{plain,label=${\bf q}$}{di,z0}
\fmf{dbl_plain,label=${\bf q}$}{z1,do}
\fmf{dbl_plain,label=${\bf q}_1$}{z0,z1}
\fmf{dashes,left}{z0,z1}
\fmfdot{z0,z1}
\end{fmfchar*}.
\end{eqnarray}

The resulting self-consistency equation for the self-energy $\Sigma({\bf q},\omega)$ is~\cite{smg}
\begin{equation}
\Sigma({\bf q},\omega) = {8 \pi \alpha^2 v^2 \over \hbar^2 L^2}
\sum_{\bf p} {e^{- \ell^2 |{\bf q}-{\bf p}|^2/2} \sin^2(\frac{\ell^2}{2} {\bf q}\wedge
{\bf p}) \over \omega + i \delta - \Sigma({\bf p},\omega)}.
\label{scba}
\end{equation}
%Here ${\hat \Pi} = (\omega + i \delta - \Sigma)^{-1}$.
Numerical solution of (\ref{scba}) in the continuum limit $N_\phi \rightarrow \infty$ finds a diffusive form for small $\hbar \omega \ell / v$ and $q \ell$:
\begin{equation}
{\hat \Pi} (q,\omega) \approx {1 \over \omega + i D_0 q^2}, 
\end{equation}
with $D_0 \approx 0.965 v \ell / \hbar$.  Note that the SCBA self-energy has a different functional form $\Sigma \sim i v q^2 \ell / \hbar$ from that of the first term in perturbation theory in $v/\omega$ ($\Sigma_1 \sim v^2 q^2 / \hbar^2 \omega$), though both are consistent with the scaling requirement
\begin{equation}
{\Sigma}({\bf q},\omega) = \omega f(v / \omega,{\bf q}).
\end{equation}

The above SCBA calculation predicts diffusion at sufficiently long times, or $\alpha=1$.
We can understand systematically what is missing in the above calculation, and hence
what causes subdiffusive scaling $\alpha < 1$, by first finding a limit in which the SCBA
result is correct.  Suppose that the counting of diagrams is modified
in the following way (in a moment the supersymmetric formalism will be introduced to
find what theory is described by these diagrams).  Let each density line carry a flavor
$i = 1,\ldots,N$, and constrain the flavor indices at each vertex to be paired as:
\begin{equation}
\begin{fmfchar*}(30,15)
\fmfstraight
\fmfbottom{di,z0,z2,z3,z1,do}
\fmf{plain,label=$i$}{di,z0}
\fmf{plain,label=$j$}{z3,z1}
\fmf{plain,label=$i$}{z1,do}
\fmf{plain,label=$j$}{z0,z2}
\fmf{dashes,left}{z0,z1}
\fmfdot{z0,z1}
\end{fmfchar*}.
\end{equation}
Simple counting then shows that a noncrossing diagram of $n$ interaction lines with fixed external indices $i=1$ has a total index degeneracy $N^n$.  Any diagram with at least one crossing has at least one constrained choice of index and hence degeneracy smaller by order $1/N$ (actually smaller by order $1/N^2$ for this particular theory) in the $N \rightarrow \infty$ limit.  If we redefine $v \rightarrow v/\sqrt{N}$ in order to keep the
effective interaction strength finite, then the large-$N$ limit of this diagrammatic sum
is exactly the SCBA result given above.

The theory which corresponds to this modified diagram expansion has the same $U(1|1)$ supersymmetry as the original $N=1$ theory, plus an additional $SU(N)$ unitary symmetry.  It will turn out that a more useful theory for understanding scaling at the
plateau transition has orthogonal rather than unitary symmetry.  Now we review the
$U(1|1)$ theory which describes the original $N=1$ problem, then consider its
different generalizations.

The density correlation function ${\hat \Pi}({\bf q},\omega)$ obtained from the
Liouvillian~\cite{smg} can be rewritten as the correlator of a complex boson field $\phi$ in a field theory with supersymmetry~\cite{efetov}.  The theory has one bosonic field $\phi$ and one fermionic field $\psi$, which appear symmetrically in the action:
\begin{eqnarray}
{\hat Pi}(q,\omega) &=& -i \int D{\bar \phi}\, D\phi\, \int D{\bar\psi}\,D \psi\,
{\bar \phi}_q \phi_q e^{-F(\omega)}, \cr
F(\omega) &=& -i \omega \int\,d{\bf q}\,
({\bar \phi}_{\bf q} \phi_{\bf q} + {\bar \psi}_{\bf q} \psi_{\bf q})
+ \cr&&\int_{1,2,3,4} f(1,2,3,4)
\Big[
{\bar \phi}_{{\bf q}_1} {\bar \phi}_{{\bf q}_2}
{\phi}_{{\bf q}_3} {\phi}_{{\bf q}_4} + \cr
&&\quad\quad 2 {\bar \psi}_{{\bf q}_1} {\bar \phi}_{{\bf q}_2}
{\phi}_{{\bf q}_3} {\psi}_{{\bf q}_4} +
{\bar \psi}_{{\bf q}_1} {\bar \psi}_{{\bf q}_2}
{\psi}_{{\bf q}_3} {\psi}_{{\bf q}_4} \Big].
\label{fieldtheory}
\end{eqnarray}
The effective interaction from disorder averaging is
\begin{eqnarray}
f(1,2,3,4) &=& {1 \over \pi} e^{-\frac{1}{2} |{\bf q}_1 - {\bf q}_4|^2}
\delta({\bf q_1} + {\bf q}_2 - {\bf q}_3 - {\bf q}_4) \cr
&&\times \sin\left(\frac{1}{2}{\bf q}_1 \wedge {\bf q}_4 \right)
\sin\left(\frac{1}{2}{\bf q}_2 \wedge {\bf q}_3 \right).
\label{nonint}
\end{eqnarray}
The effect of the supersymmetry between the bosonic and fermionic fields can be understood in terms of the Feynman diagrams defined above: its effect is to
cancel all diagrams for ${\hat \Pi}({\bf q},\omega)$ except those which contain
a single bosonic line running through the diagram.  Such supersymmetry is a general
feature of the field theories which describe noninteracting quantum-mechanical electrons,
but the Liouvillian theory (\ref{fieldtheory}) is superficially quite different from conventional $\sigma$-model descriptions, due to the nonlocal interaction (\ref{nonint}).

The class of large-$N$ generalizations introduced in~\ref{mooresinova} are of the form (note that here the definition of $c_i$ is slightly modified to reflect standard usage: now ${c_i}^2$ rather than $c_i$ is the coefficient of a term in the action)
\begin{eqnarray}
F_{\rm gen}(\omega)&=&- i \omega \sum_{i=1}^N \int d{\bf q}\,
({\bar \phi}^i_{\bf q} \phi^i_{\bf q} + {\bar \psi}^i_{\bf q} \psi^i_{\bf q}) + F_{\rm int},\cr
F_{\rm int} &=&\sum_{i,j=1}^N \int\,d{\bf q}_1\,d{\bf q}_2 {f(1,2,3,4) \over N}
\Big[ {c_1}^2
({\bar \phi}^i_{{\bf q}_1} {\bar \phi}^j_{{\bf q}_2}
{\phi}^i_{{\bf q}_3} {\phi}^j_{{\bf q}_4} +
 2 {\bar \psi}^i_{{\bf q}_1} {\bar \phi}^j_{{\bf q}_2}
{\phi}^i_{{\bf q}_3} {\psi}^j_{{\bf q}_4} +
{\bar \psi}^i_{{\bf q}_1} {\bar \psi}^j_{{\bf q}_2}
{\psi}^i_{{\bf q}_3} {\psi}^j_{{\bf q}_4} ) \cr
&&\quad+{c_2}^2 ({\bar \phi}^i_{{\bf q}_1} {\bar \phi}^j_{{\bf q}_2}
{\phi}^j_{{\bf q}_3} {\phi}^i_{{\bf q}_4} +
 2 {\bar \psi}^i_{{\bf q}_1} {\bar \phi}^j_{{\bf q}_2}
{\phi}^j_{{\bf q}_3} {\psi}^i_{{\bf q}_4} +
{\bar \psi}^i_{{\bf q}_1} {\bar \psi}^j_{{\bf q}_2}
{\psi}^j_{{\bf q}_3} {\psi}^i_{{\bf q}_4} ) \cr
&&\quad+{c_3}^2 ({\bar \phi}^i_{{\bf q}_1} {\bar \phi}^i_{{\bf q}_2}
{\phi}^j_{{\bf q}_3} {\phi}^j_{{\bf q}_4} +
 2 {\bar \psi}^i_{{\bf q}_1} {\bar \phi}^i_{{\bf q}_2}
{\phi}^j_{{\bf q}_3} {\psi}^j_{{\bf q}_4} +
{\bar \psi}^i_{{\bf q}_1} {\bar \psi}^i_{{\bf q}_2}
{\psi}^j_{{\bf q}_3} {\psi}^j_{{\bf q}_4} )
\Big] \cr
&=&{c_1}^2
\begin{fmfchar*}(20,10)
\fmfstraight
\fmfbottom{di,z0,z2,z3,z1,do}
\fmf{plain,label=$i$}{di,z0}
\fmf{plain,label=$j$}{z3,z1}
\fmf{plain,label=$i$}{z1,do}
\fmf{plain,label=$j$}{z0,z2}
\fmf{dashes,left}{z0,z1}
\fmfdot{z0,z1}
\end{fmfchar*}
+{c_2}^2
\begin{fmfchar*}(20,10)
\fmfstraight
\fmfbottom{di,z0,z2,z3,z1,do}
\fmf{plain,label=$i$}{di,z0}
\fmf{plain,label=$j$}{z3,z1}
\fmf{plain,label=$j$}{z1,do}
\fmf{plain,label=$i$}{z0,z2}
\fmf{dashes,left}{z0,z1}
\fmfdot{z0,z1}
\end{fmfchar*}
+{c_3}^2
\begin{fmfchar*}(20,10)
\fmfstraight
\fmfbottom{di,z0,z2,z3,z1,do}
\fmf{plain,label=$i$}{di,z0}
\fmf{plain,label=$i$}{z3,z1}
\fmf{plain,label=$j$}{z1,do}
\fmf{plain,label=$j$}{z0,z2}
\fmf{dashes,left}{z0,z1}
\fmfdot{z0,z1}
\end{fmfchar*}.
\end{eqnarray}
The physical content of these generalizations will become clearer in Section V, when we discuss the noninteracting boson problems that generate them upon disorder averaging.
In order to reduce to the original theory at $N=1$, we impose ${c_1}^2 + {c_2}^2 + {c_3}^2 = 1$.  The properties of the generalizations are reviewed only briefly here, since more details are in~\cite{mooresinova}.  With ${c_3}^2=1, {c_1}^2 = {c_2}^2 = 0$, the only diagrams which survive have just one electron flavor running all the way through.  In this case the theory for any $N$ is just equivalent to the $N=1$ theory, and has a full $U(N|N)$ supersymmetry, as each index can be rotated independently (this equivalence for the fully symmetric theory is discussed more generally in~\cite{readsaleur}).

Other values of the $c_i$ give theories which for finite $N$ are physically different from
the $N=1$ theory, and are analytically solvable in the large-$N$ limit.  The mean-field theory of~\ref{smg} corresponds to the large-$N$ limit of the ${c_1}^2=1$ theory,
which breaks the $U(N|N)$ symmetry down to $SU(N) \times U(1|1)$.  This theory was discussed earlier in this section.  The ${c_1}^2 = {c_2}^2 = 1/2$ theory has a logarithmically divergent $1/N$ correction~\cite{mooresinova}, whose physical explanation is discussed in the following section.  For this theory the symmetry is further broken to $SO(N) \times U(1|1)$.  The $N\rightarrow \infty$ limit is still diffusive, although the diffusivity is reduced by a factor $\sqrt{2}$ from the unitary case.

The physical meaning of the $N>1$ generalizations is that they correspond to problems of $N$ different flavors of bosons moving in random potentials.  The bosons are noninteracting and boson number is conserved, as expected since the theories remain supersymmetric.  The $U(N|N)$ case corresponds to random potentials which never scatter bosons of one flavor into bosons of another flavor: then these are trivially equivalent to the $U(1|1)$ case since the flavors are independent.
The difference between the orthogonal $SO(N) \times U(1|1)$ and unitary $SU(N) \times U(1|1)$ generalizations, for example, is in the correlations between the random potentials.  The numerical results in Section V confirm that these theories are nontrivial generalizations with physics that interpolates between the $N=1$ and $N=\infty$ cases.

\section{Conservation laws and polygon identities}

The density operators, which are the fundamental quantities in the Liouvillian approach, must satisfy a very large number of conservation laws.  If there are $N_\phi$ states in the LLL, there are $N_\phi^2$ different density operators, which cannot be assigned values independently: only some values of the density operators correspond to actual LLL states.  A simple example is that there is no one-electron LLL state with uniform density (i.e., $\langle \psi | \rho_0 \psi \rangle = 1$, $\langle \psi | \rho_{\bf q} | \psi \rangle = 0$ for ${\bf q} \not = 0$).  The constraints on allowed values of density operators will turn out to be important in understanding how approximations to the exact Liouvillian theory.

A convenient way to represent the constraints on density operators is in terms of geometric ``polygon identities.''  Essentially the Liouvillian time evolution must conserve a huge number of quantities, for any realization of disorder, in order that the density operators always correspond to an allowed LLL state.  The first two conservation laws are relatively simple and are preserved by the mean-field and large-$N$ theories, but higher conservation laws contain geometric phase factors and are not always conserved.

The first conserved quantity is simply the particle number: $C_0 = \rho_0 = 1$.
The second is the ``chop'' $C_1$: for all LLL states,
\begin{equation}
C_1 \equiv \sum_{\bf q} \langle \psi | \tau_{\bf q} |\psi \rangle \langle \psi | \tau_{\bf -q}
| \psi \rangle = 
\sum_{\bf q} |\langle \psi | \tau_{\bf q} | \psi \rangle|^2 = N_\phi.
\end{equation}
Here the sum is extended over the $N_\phi^2$ different momenta in the LLL.  This conservation law means that all single-particle LLL states have a certain amount of variation of the density, in addition to having fixed total density.  It prohibits the uniform density state mentioned above, since then $C_1 = 1$.
Note that the expectation values make this different from the trivial statement $\sum_{\bf q} \tau_{\bf q} \tau_{- \bf q} = \sum_{\bf q} = N_\phi^2.$
These conservation laws clearly have consequences for correlation functions of the Liouvillian theory, as explained in more detail below.  Now it is shown that $C_0$ and $C_1$ are just the first two of a hierarchy of independent constraints on the density operators; each constraint induces additional restrictions on correlation functions of the Liouvillian.

\begin{figure}
\begin{center}
\includegraphics[width=2.0in]{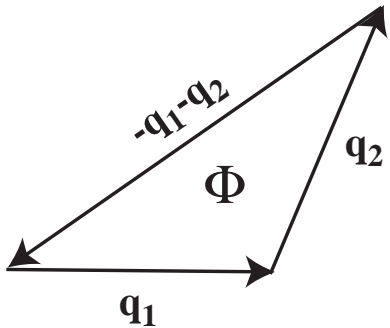}
\end{center}
\caption{The conserved quantity $C_2$ is defined as a sum over triangles such as the above, weighted by the phase factor $e^{i \Phi}$.}
\label{figtrig}
\end{figure}

Consider translating an electron around a triangle (Fig.~\ref{figtrig}) by successive action of $\tau_{{\bf q}_1}$, $\tau_{{\bf q}_2}$, and $\tau_{-{\bf q}_1-{\bf q}_2}$.  In doing so the electron picks up a phase factor set by the signed area of the triangle,
\begin{equation}
\exp(i \Phi) = \exp(i \ell^2 {\bf q}_1 \wedge {\bf q}_2 / 2).
\end{equation}
Note that $C_1$ can be thought of as summing over $2$-gons, where the electron is first translated by a vector $\ell^2 {\bf q} \times {\bf \hat z}$ and then in reverse.  Such a $2$-gon encloses no area, so $\Phi = 0$.  Similarly $C_0$ can be thought of as summing over the $1$-gon of zero momentum.  The generalization of $C_0$ and $C_1$ to triangles is
\begin{eqnarray}
C_2 \equiv \sum_{{\bf q}_1,{\bf q}_2} \langle \tau_{{\bf q}_1} \rangle
\langle \tau_{{\bf q}_1} \rangle \langle \tau_{-{\bf q}_1 - {\bf q}_2} \rangle
e^{-i \ell^2 {\bf q}_1 \wedge {\bf q}_2 / 2} = N_\phi^2.
\end{eqnarray}
Here all three expectation values are taken in one LLL state.  Note that periodic or antiperiodic boundary conditions must be imposed consistently in this sum when ${\bf q}_1 +{\bf q}_2$ lies outside the original set of $N_\phi^2$ momenta, in order to preserve the correct commutation relations.

A quick proof that these quantities take fixed values for all states in the LLL starts from the lemma that for any state $|\psi \rangle$, the sum over its ${N_\phi}^2$ translations,
properly normalized, forms a resolution of the identity:
\begin{equation}
\sum_{\bf q} {\tau_{\bf q} |\psi \rangle \langle \psi | \tau_{\bf -q} \over N_\phi}= {\bf 1}.
\label{resident}
\end{equation}
This can be shown explicitly using e.g. the Landau basis of LLL states, in which
\begin{equation}
\sum_{\bf q} \langle \alpha | \tau_{\bf q} |\gamma \rangle \langle \epsilon | \tau_{\bf -q} | \beta \rangle = N_\phi \delta_{\alpha \beta} \delta_{\gamma \epsilon}.
\end{equation}
Then the commutation relations and (\ref{resident}) can be used to prove by induction the polygon identity
\begin{equation}
C_n \equiv \sum_{{\bf q}_i, 1 \leq i \leq n} \left( \prod_i \langle \tau_{{\bf q}_i} \rangle \right) \langle \tau_{- \sum_i {\bf q}_i} \rangle e^{-i \Phi} = N_\phi^n.
\end{equation}
Here $\Phi$ is the signed flux through the area of the polygon formed by the ${\bf q}_i$.

These conservation laws imply that the $N$-body density correlation functions have
singularities at zero total momentum and small frequency.  Conservation of particle number $C_0$ implies, of course, that the one-body density correlation introduced previously has a singularity at zero momentum:
\begin{equation}
{\hat \Pi}({\bf q}=0,\omega) = {1 \over \omega + i \delta}.
\end{equation}
Conservation of the chop $C_1$ implies a singularity in the two-body correlation function
at zero total momentum, for any initial momentum, but summed over final momentum:
\begin{equation}
\sum_{{\bf q}_1}
{\hat \Pi}^2({\bf q},-{\bf q}, {\bf q}_1, -{\bf q}_1, \omega) = {1 \over \omega + i \delta}.
\end{equation}
Similar singularities exist in the three-body and higher correlation functions because of the higher conservation laws mentioned above, although now the geometric factor $e^{i \Phi}$ must be included in the sum over correlation functions.

The singularity of the two-body correlation function $C_1$ is responsible for the logarithmic corrections found at order $1/N$, where $N$ is the number of density flavors, in the following way.  First, the singularity becomes a diffusive pole for small total momentum.  This is the diffusive pole that appears in the diverging ladder sum which enters into the maximally crossed diagrams for the one-body correlation function.
These diagrams appear at order $1/N$, where $N$ is the number of density flavors, causing the logarithimic corrections discussed in~\cite{mooresinova} and the preceding section.  The conservation of $C_1$ even in the large-$N$ generalizations (see below) explains why there is a diffusive pole from ladders with two density lines, corresponding to {\it four} fermionic lines rather than the two lines in the diagrams leading to the weak-localization singularity.  Hence the logarithmic divergence at order $1/N$ has a different physical origin than the logarithmic divergence in weak localization.

An important difference between $C_0$ and $C_1$ on one hand, and the higher conservation laws on the other, is that the $N>1$ generalizations still preserve analogues of $C_0$ and $C_1$:
\begin{eqnarray}
C_0^i &=& \langle \tau^i_0 \rangle, \cr
C_1 &=& \sum_{i,{\bf q}} \langle \tau^i_{{\bf q}} \rangle \langle \tau^i_{-{\bf q}} \rangle
\end{eqnarray}
are all conserved, where $i = 1, \ldots, N$.  There is no such generalization for the quantities $C_n$, $n\geq 2$.  The next section shows how numerical simulations can be used to study the $N>1$ generalizations in the same way as the well understood $N=1$ case.  The numerical simulations suggest that preserving the entire hierarchy of conservation laws $C_n$ is important for obtaining the true anomalous (i.e., not diffusive) scaling.

\section{Numerical results}

Section \ref{review} derived the Liouvillian equation of motion for the magnetic translation operators $\tau_q$,
\begin{equation}
{\dot \tau_{\bf q}}(t) = -i \sum_{\bf q^\prime} {\cal L}_{\bf q q^\prime}
\tau_{\bf q^\prime}(t).
\label{eom2}
\end{equation}
This evolution is in the Heisenberg representation: (\ref{eom2}) also describes
the evolution of the expectation values $\langle \tau_q(t) \rangle$ in a specific state, from the initial values $\langle \tau_q(0) \rangle$.

Recall that the Liouvillian ${\cal L}$ is an ${N_\phi}^2 \times {N_\phi}^2$ matrix for a torus of $N_\phi$ states in the LLL, while the Hamiltonian is only of size $N_\phi \times N_\phi$.  Hence direct numerical calculation of the Liouvillian equation (\ref{eom2}) is grossly inefficient compared to evolution of the ordinary Schr{\"o}dinger equation $-i \hbar {\dot \psi} = H \psi$.
Previous numerical studies of the Liouvillian approach~\cite{smg,boldyrev,sandler} have therefore used the Schr{\"o}dinger equation to evolve $\psi(t)$: then the expectation
values of the translation operators, and hence densities, are just
\begin{equation}
\langle \tau_{\bf q}(t)\rangle = \langle \psi(t) | \tau_{\bf q} | \psi(t)\rangle.
\end{equation}
So the evolution of the ${N_\phi}^2 \times {N_\phi}^2$-dimensional Liouvillian problem
can be found exactly from the reduced Hamiltonian problem of size $N_\phi \times N_\phi$.  The Liouvillian {\cal L} is therefore quite redundant: for example, its ${N_\phi}^2$ eigenvalues are just all the differences of pairs of eigenvalues of $H$.

Now let us consider a generalization of the Liouvillian equation of motion (\ref{eom}) to $N$ flavors of translation operators $\tau^i_{\bf q}$, and associated density operators
$\rho^i_{\bf q} \equiv e^{-q^2 \ell^2/4} \tau_{\bf q}$:
\begin{equation}
{\dot \tau^i_{\bf q}}(t) = -i \sum_{\bf q^\prime} {\cal L}^{ij}_{\bf q q^\prime}
\tau^j_{\bf q^\prime}(t).
\label{neom}
\end{equation}
In a moment we will define the generalized Liouvillian ${\cal L}^{ij}$.  Before there were
${N_\phi}^2$ different operators $\tau_{\bf q}$ which had to be evolved; now there are $N {N_\phi}^2$ operators.  This simple generalization to $N$ flavors will turn out to
have far-reaching consequences for the behavior of (\ref{neom}).  The key difference is that for $N>1$ there is no underlying Hamiltonian of which ${\cal L}^{ij}$ is
the Liouvillian; this explains how the $N>1$ generalizations of the Liouvillian differ
from previous work on large-$N$ generalized Hamiltonians.

The generalized Liouvillian ${\cal L}^{ij}$ for $N>1$ differs from the original ${\cal L}$ only in having multiple disorder potentials $v^{ij}$:
\begin{equation}
{\cal L}_{\bf q q^\prime} = {2 i \over \hbar} v^{ij}({\bf q} - {\bf q}^\prime)
\sin(\frac{\ell^2 {\bf q} \wedge {\bf q}^\prime}{2})
e^{-\frac{\ell^2}{4} |{\bf q} - {\bf q}^\prime|^2}.
\end{equation}
There are several possible constraints on the potentials $v^{ij}$.  For instance if $v^{ij} = 0$ for $i \not = j$, then ``flavor'' is conserved and clearly the dynamics are the same as
for $N=1$.  We will concentrate on the case of independent Gaussian random variables:
\begin{eqnarray}
v^{ij} (\bf q) &=& v^{ij}({\bf -q})^* = v^{ji}({\bf q})\cr
\langle \langle v^{ij} ({\bf q}) \rangle \rangle &=& 0 \cr
\langle \langle v^{ij} ({\bf q}) v^{kl} ({\bf q^\prime})\rangle \rangle &=& 
{2 \pi v^2 \over 2 L^2} \delta_{{\bf q}+{\bf q^\prime}} (\delta_{ik} \delta_{jl} +
\delta_{il} \delta{jk}).
\end{eqnarray}
This will turn out to yield the orthogonal generalizations mentioned in Section III.  The unitary generalizations can be obtained by a weaker constraint on the random potentials: instead of $v^{ij}({\bf q}) = v^{ij}(-{\bf q})^*$, one imposes only 
$v^{ij}({\bf q}) = v^{ji}(-{\bf q})^*$.

With this choice of potentials, the equations of motion (\ref{neom}) for $N>1$ do not have a reduced representation in the same way as the $N=1$ problem: the generalized Liouvillian for $N>1$ is not in general the Liouvillian matrix of any smaller matrix.  However,
note that (\ref{neom}) is still the equation of motion of a noninteracting disordered
problem, and that the Hilbert space is only polynomial in the system size, rather than
exponential as for an interacting problem.  Another important observation is that the $N>1$ problems (\ref{neom}), when disorder-averaged, become exactly the supersymmetric $SO(N) \times U(1|1)$ theories discussed in Ref. \cite{mooresinova} and
section III.  So the meaning of the supersymmetry of those theories is
now clear: they describe $N$ flavors of noninteracting real bosons, which for $N=1$ are density operators of noninteracting fermions.  For $N>1$, it is shown below that
the bosons cannot be interpreted as density operators of some underlying fermions.

Now we return to the original $N=1$ problem and introduce the numerical methods
which will later be applied for $N>1$.  The system is on a torus pierced by $N_\phi$ flux
quanta: the Landau basis for the $N_\phi$ states of the LLL is
\begin{equation}
\psi_k(x,y) = {1 \over \sqrt{2 \pi L \ell}} \sum_m e^{2 \pi i k (y/L+m)}
e^{-(x-m L-k L/N_\phi)^2 /2\ell^2}.
\label{landau}
\end{equation}
Here $-L/2 \leq x,y \leq L/2$.  For definiteness fix $N_\phi$ odd: then $N_\phi =
2 Q + 1$ and $k=-Q,\ldots,Q$ in (\ref{landau}).  There are $N_\phi$ independent values of momentum in one direction, spaced by ${2 \pi \over L} = \sqrt{2 \pi / \ell^2 N_\phi}$: $\ell q_x = n_x \sqrt{2 \pi / N_\phi}$, with $n_x = -Q,\ldots, 0, \ldots,
Q$.

Now imagine starting the system at $t=0$ in the state $\psi_0$, which is localized along
the line $x=0$.  Over time, the wavefunction will spread out under the influence of the
random potential; this spreading can be studied numerically to estimate $\alpha = 1-(2 \nu)^{-1}$..
The Hamiltonian for a given random potential $v({\bf q})$ is
\begin{equation}
H = \sum_{\bf q} v(-{\bf q}) {\hat \rho}_{\bf q} = \sum_{\bf q} v(-{\bf q})
\tau_{\bf q} e^{- q^2 \ell^2 / 4}.
\label{fullham}
\end{equation}
The matrix elements of the translation operators in the basis (\ref{landau}) are~\cite{boldyrev}
\begin{equation}
\langle j | \tau_{n_x,n_y} | k \rangle =
e^{i \pi (n_x n_y+2 j n_x} \delta(j-k \equiv n_y\ {\rm mod}\ N_\phi).
\end{equation}

The approach used here starts with a localized state and evolves it under the Hamiltonian (\ref{fullham}) to calculate the mean squared displacement over time.
This approach is almost as efficient as direct diagonalization of the Hamiltonian
for the single-flavor problem $N=1$, but its real numerical advantage is for the large-$N$ generalizations introduced below, where direct diagonalization is unavailable.

To calculate displacements, we use the expression for $\langle x^2 \rangle$ in the LLL for a torus with $N_\phi$ flux quanta:
\begin{equation}
\langle x^2 \rangle \ell^{-2} = {\pi N_\phi \tau_{0,0} \over 6} + N_\phi
\sum_{n=1}^\infty {e^{-\pi n^2 / 2 N_\phi} (-1)^n (\tau_{n,0} +
\tau_{-n,0})\over
\pi n^2}.
\label{sep}
\end{equation}
This form may be regarded as a discrete approximation to the second derivative with regard to $q_x$ which appears in the continuum limit (\ref{characteristic}). 
For the state $\psi_0$ localized around $x=0$, we have $\tau_{n,0} = 1$
and $\langle x^2 \rangle \rightarrow \ell^2/2$ as $N_\phi \rightarrow \infty$.
The rotationally invariant state of lowest angular momentum $\psi_r$ as $N_\phi \rightarrow \infty$
has $\langle x^2 \rangle = \langle y^2 \rangle = \ell^2$.

We define a quantity $D_{\rm est}(t)$ which estimates the diffusion constant over the interval $(t/2,t)$:
\begin{equation}
D_{\rm est}(t) \equiv {\langle \langle x^2(t) \rangle \rangle - \langle \langle x^2(t/2) \rangle \rangle \over t/2}.
\label{dest}
\end{equation}
The motivation for using the interval $(t/2,t)$ is to weaken the effect of the ordered initial condition at $t=0$, while still averaging over a long interval in time to decrease the
statistical fluctuations.

\begin{figure}
\begin{center}
\includegraphics[width=3.0in]{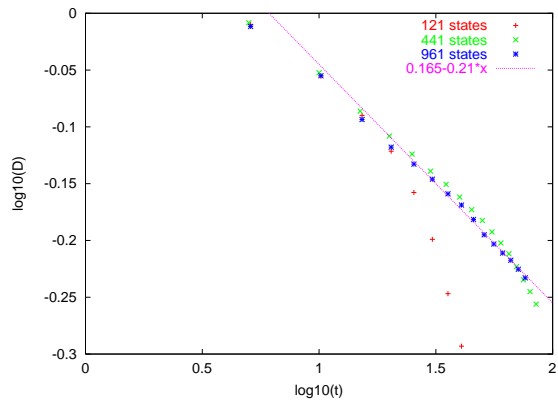}
\end{center}
\caption{The decrease of the effective diffusion strength $D_{\rm est}(t)$ with time
for three different system sizes.  Time is measured in dimensionless units $v t /\hbar \ell$, and the diffusion strength is normalized to the mean-field value.  The solid line shows $D \propto t^{-0.21}$, the value quoted in the text.
}
\label{figtwo}
\end{figure}

Results for $D_{\rm est}(t)$ for $N=1$ are shown for system sizes from $N_\phi=81$ to $N_\phi = 441$ in Fig.~\ref{figtwo}.  After a short initial period, $D_{\rm est}$ is observed
to decline as a power-law until the density begins to reach the boundary of the system:
\begin{equation}
D_{\rm est}(t) \sim t^{-0.21 \pm 0.01} \Rightarrow \alpha \approx 0.79 \pm 0.01.
\end{equation}
This estimate of $\alpha$ corresponds to $2.27 \leq \nu \leq 2.5$, consistent with
the accepted value~\cite{huckestein}.  Even for fairly small system sizes $N \geq 200$ there is a significant region where power-law scaling is observed.  Results on the $N=1$ system can also be used to estimate the scale of finite-size effects for $N>1$.

The main lesson of the above is that it is possible to estimate $\alpha$ from quite small
systems even without diagonalizing the Hamiltonian completely.  It is sufficient to be
able to observe the evolution of one initial state for many different disorder realizations.  Such evolution is numerically fast even for the $N>1$ problem, where direct diagonalization of the sparse ${N_\phi}^2 \times {N_\phi}^2$ Liouvillian matrices would
require a very large amount of memory.  The results in~\ref{figtwo} are similar to those reported for one system size in~\cite{boldyrev}, but differ in that we have used the full Hamiltonian (\ref{fullham}) without truncation, and used $D_{\rm est}$ (\ref{dest}) to measure displacements.  There is also a recent detailed study of the displacement scaling in long-range correlated potentials~\cite{sandler}.

Now the $N>1$ generalizations (\ref{neom}) can be considered using the same technique.
The initial values of the $\tau$ operators are
\begin{equation}
\tau^i_{k,l}(0) = \delta_{i1}\delta_{0l}. 
\label{initcond}
\end{equation}
Then after time $t$, the disorder-averaged correlation function ${\hat \Pi}({\bf q},t)$ is is estimated, following (\ref{piest}), as
\begin{equation}
{\hat \Pi}(2 \pi k {\bf \hat x}/L,t) = \langle \langle \tau^1_{k,0} \rangle \rangle.
\end{equation}

In order to calculate $\alpha$, which is determined by the second derivative of
${\hat \Pi}({\bf q},t)$ at ${\bf q}=0$, we recall that the formula for $\langle \langle x^2
\rangle \rangle$ on the torus (\ref{sep}) is just an approximation to this derivative in
a bounded system.  That is, $\alpha$ is estimated through
\begin{equation}
s(t) \equiv {\pi N_\phi \over 6} + N_\phi
\sum_{n=1}^\infty {e^{-\pi n^2 / 2 N_\phi} (-1)^n (\tau^1_{n,0}(t) +
\tau^1_{-n,0}(t))\over
\pi n^2} \sim t^\alpha.
\label{nsep}
\end{equation}
For large $N_\phi$, $s(0) = \frac{1}{2}$ for the initial condition (\ref{initcond}).  Note that it no longer makes sense to speak of a wavefunction $\psi(t)$
and displacement $\langle \psi | x^2 | \psi \rangle$.  The densities $\tau^i_{\rm q}(t)$ cannot generally be obtained as expectation values in any state $\psi^i(t)$ via $\langle \psi^i(t) | \tau_{\rm q} | \psi^i(t)\rangle$.

In order to compare $N=2$ results with $N=1$ results, the disorder strength $v$ of the $N=2$ problem must be scaled by $1/\sqrt{2}$ relative to the strength of the $N=1$
problem.  This is an example of the $1/\sqrt{N}$ rescaling required to obtain a finite theory as $N \rightarrow \infty$ in section III.  Now the generalized Liouvillian equations of motion (\ref{neom}) are integrated numerically, and the
conservation laws of section IV can be used used to check the accuracy of this integration.

Another check on the $N=2$ numerics is the following test, which also clarifies the
difference between the $N=2$ and $N=1$ problems.  There are three independent disorder potentials at each momentum ${\bf q}$: $v^{11}({\bf q})$, $v^{12}({\bf q})$, and $v^{22}({\bf q})$.  With the added constraint $v^{11} = v^{22}$, there are ``odd'' and ``even'' densities which evolve independently: $\tau^\pm({\bf q}) \equiv
(\tau^{11}_{\bf q} \pm \tau^{22}_{\bf q})/\sqrt{2}$ evolve like the $N=1$ problem
in the disorder potentials $v^{\pm} \equiv (v^{11} \pm v^{12})/\sqrt{2}$.  Another way to express this change is that the theory's symmetry is enlarged from $SO(2) \times U(1|1)$ to $U(1|1) \times U(1|1)$ by the extra constraint.  Numerics confirm that
this constrained $N=2$ theory has $D_{\rm est}(t)$ equal to that of the $N=1$ theory,
and quite different from that of the true $N=2$ theory for sufficiently long times.

We now summarize briefly the main conclusions.  The large-$N$ generalizations of the Liouvillian theory of the quantum Hall effect are supersymmetric theories that describe problems of bosons moving in a disorder potential.  They preserve some but not all of an infinite hierarchy of conserved quantities in the Liouvillian theory, reflecting the interdependence of the momentum operators.  The $N>1$ generalizations interpolate between the $N=1$ quantum Hall case and the mean-field theory at $N=\infty$, and can be studied numerically using Monte Carlo simulations.  We find that an initially localized wave packet spreads diffusively for sufficiently long times, although the diffusion constant is significantly reduced from the mean-field value.  This study confirms that large-$N$ generalization of the Liouvillian formalism is a controlled expansion to calculate some properties of the problem, but suggests that the full hierarchy of conserved quantities is required to obtain the complete quantum Hall plateau transition.

\begin{figure}
\begin{center}
\includegraphics[width=3.0in]{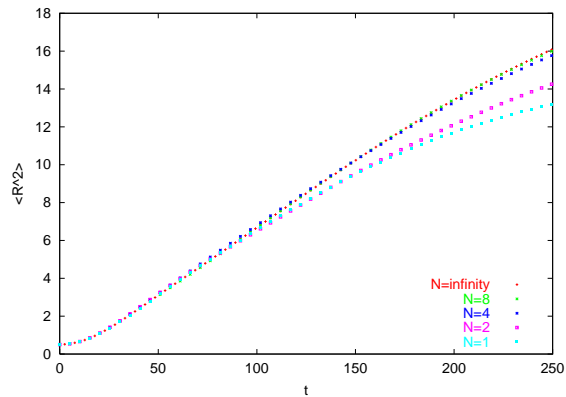}
\end{center}
\caption{The evolution of $\langle x^2 \rangle$ in units of $\ell^2$ for a finite torus system of $N_\phi = 49$ flux quanta, for different values of the flavor index $N$.  The
$N=\infty$ result is obtained from analytic solution of the mean-field equations; the $N=1$ solution from direct evolution under the Hamiltonian; and $N=2,3,4,8$ from
direct evolution under the generalized Liouvillian.
}
\label{figdiffn}
\end{figure}

\begin{figure}
\begin{center}
\includegraphics[width=3.0in]{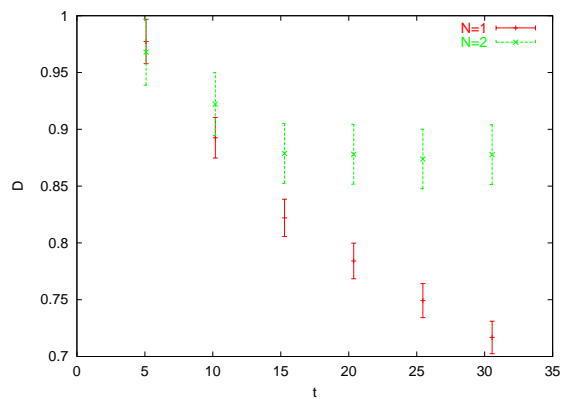}
\end{center}
\caption{The evolution of $D_{\rm est}(t)$ in units of $D_{N=\infty}$ for a finite torus system of $N_\phi = 225$ flux quanta, for $N=1$ and $N=2$.  Note that while the $N=1$ diffusion constant continues to decrease, the $N=2$ diffusion constant initially decreases but at long times begins to level off.
}
\label{figcomp}
\end{figure}

The spreading of an initially localized state with time is shown for various $N$ in Fig.~\ref{figdiffn}.  As expected, the finite-$N$ theories interpolate between the physical quantum Hall case ($N=1$) and the mean-field theory at $N=\infty$.  For this set of numerics, the system size (49 flux quanta) is too small in order to determine whether the $N>1$ theories are diffusive ($\alpha=1$) or show anomalous scaling: the wave packet reaches the system boundary before the asymptotic scaling regime is reached.

For the $N=2$ case, system sizes up to 225 flux quanta can be considered in order to study whether the evolution of a wave packet is diffusive at long times.  Fig.~\ref{noscal} shows the time-dependent diffusion strength $D(t)$ extracted from simulations for $N=1$ and 225 states in the LLL, compared to $D(t)$ for $N=2$ for the same system size (now the Liouvillian matrix is of dimension $2 (225)^2$).
Note that for the longest accessible times, the $N=2$ problem appears to show $D(t)$ leveling off, so $\langle R^2 \rangle \propto t$, rather than $D(t)$ falling to zero as a power-law.  We note in passing that the time evolution of a state under such large Liouvillian matrices can be obtained relatively rapidly because the Liouvillian matrices are quite sparse; finding all eigenvectors and eigenvalues, which is quite simple for the $N=1$ Hamiltonian of 225 states, would require a great deal of computer memory and time for the Liouvillian of the corresponding $N=2$ problem.

{\it Note added}: A preprint by V. Oganesyan, J. Chalker, and S. Sondhi ({\tt cond-mat/0212232}) posted soon after this manuscript's submission contains a similar conclusion about the behavior of the generalized Liouvillian, reached using different techniques.

\section{Acknowledgments}

The author thanks A. Green, I. Gruzberg, V. Oganesyan, N. Read, and N. Sandler for useful comments and correspondence.  A grant of computer time was provided by the NERSC facility of Lawrence Berkeley National Laboratory.

\end{fmffile}
\bibliography{newliou}
\end{document}